\documentclass[prb,showpacs,twocolumn,aps]{revtex4}
\usepackage{graphicx}
\usepackage{epsfig}

\begin{document}

\title{Localizable Entanglement in Antiferromagnetic Spin Chains}
\author{B.-Q.\ Jin and V.E.\ Korepin}
\affiliation{C.N.\ Yang Institute for Theoretical Physics,
 State University of New York at Stony Brook, Stony Brook, NY 11794-3840}
\date{\today}

\begin{abstract}

 Anti-ferromagnetic spin chains play an important role in condensed matter
 and statistical mechanics. Recently XXX spin  chain was
discussed in relation to  information theory.
Here we consider localizable entanglement. It is how much entanglement can
be localized  on two spins  by  performing local
measurements on  other individual spins (in a system of many
interacting spins). We consider a ground state of
antiferromagnetic spin chain.
We  study  localizable  entanglement [represented by concurrence] between two
spins. It is a function  of the distance.  We start with
isotropic  spin chain.
 Then we study effects of anisotropy and magnetic field.
We conclude that anisotropy increases the localizable entanglement.
We discovered high sensitivity to a magnetic field in cases of high symmetry.
We also evaluated concurrence of these two spins before the
measurement to illustrate that the measurement raises the concurrence.

\end{abstract}


\maketitle

Spin chains play an important role in solid state
physics\cite{e1,e2,tn,levit,et,st}. For example, inelastic neutron
scattering on $\mbox{SrCuO}_2$\cite{zal} can be explained by spin
1/2 Heisenberg chain\cite{es}.
 Spin chains are also important for
information theory. Recently, many-body systems attract lots of
attention in the field of quantum information
\cite{fazio,jk,bose,vpc}. Special attention has been paid to the
entanglement in these systems. Entanglement plays the main role as
a physical resource for quantum information and quantum
computation.  There is a lot in common between quantum statistical
mechanics and quantum information theory. The role of phase
transitions for quantum information was emphasized in
Ref.~\onlinecite{fazio}. Most direct relation between correlation
functions  and entanglement was discovered  by F.\ Verstraete, M.\
Popp and J.I.\ Cirac [we shall use abbreviation VPC] in
Ref.~\onlinecite{vpc}. They found that correlation function
provides a bound for localizable entanglement (LE).

The LE of two spins is defined as the maximal amount of
entanglement that can be localized on  two marked spins on average
by doing local measurements on the rest of the spins [assisting
spins]. Here we assume that we consider a pure state $|\phi
\rangle$ of all these spins. The LE has an operational meaning
applicable to situations  in which one would like to concentrate
as much entanglement as possible between two particular particles
out of multi-particle entangled state. Good examples are quantum
repeater \cite{Br98} and  spinotronics \cite{loss}. Let us
consider an example of $N$ qubits in GHZ state:
\begin{equation}
|GHZ>= {\frac{1}{\sqrt{2}}}(|00\ldots 0 \rangle  +  |11\ldots 1
\rangle )\ .
\end{equation}
We can measure  assisting qubits in $|\pm \rangle $ basis. This
will force two marked spins
 into a Bell state [maximally entangled state of two qubits].

Let us proceed to the  formal definition for localizable
entanglement $E_{ij}$ between two  spins marked by  $i$  and $j$.
Consider  a pure state of $N$ spins $|\phi \rangle $ [it is
normalized $\langle \phi | \phi \rangle = 1$]. Every measurement basis
specifies an ensemble of pure states  ${\mathcal{E}}=\{p_s, |
\psi_s \rangle \}$. The index $s$  enumerates different
measurement outcomes. It runs through
 $2^{(N-2)}$ values. Here $ |\psi_s \rangle $  is a two-spin state
after the measurement and $p_s$ is its probability. The LE is defined  as
\begin{equation}
 E_{ij}= \max_{\mathcal{E}} \sum_s p_s E(|\psi_s \rangle)\ .
\end{equation}
Here $ E(|\psi_s \rangle)$  is the entanglement of $|\psi_s
\rangle $, characterized by   concurrence.
The  concurrence  $C$  was suggested by  W.K. Wooters\cite{wot} as a
measure of entanglement
 \footnote{Entropy of entanglement is
equal to $H(\{ 1+{\sqrt{1-C^2}}\}/2)$,  here $H(x) $ is Shanon
entropy. }. By definition it is  $0\leq C\leq 1$.  VPC  noticed that it is in
particular convenient measure for LE. It is important for us that
 concurrence for
two qubits state $ | \phi\rangle = a |00\rangle +b|01\rangle
+c|10\rangle +d |11\rangle $  coincides with maximum correlation
$C(|\phi \rangle ) = 2|ad-bc|$.

It is difficult to calculate  LE. Instead,  VPC found
 bounds for LE. The upper bound comes out of considering a
global [joint] measurement on all assisting spins. It can be
related to the entanglement of assistance, which is the maximum
 entanglement over all possible states of $N$ spins
consistent with the density matrix of two marked spins. It was  introduced
by D.P. DiVincenzo, C.A. Fuchs, H.Mabuchi, J.A.Smolin, A.Thapliyal
and A.Uhlmann  \cite{div}. A simple formula for
entanglement of assistance was found in Ref.~\onlinecite{enk}.
Let us denote the density matrix of two marked spins by $\rho_{ij}$.
Matrix $X$ is a square root of the density matrix: $\rho_{ij}=
XX^\dagger$. The entanglement of assistance  measured by
concurrence is given by the trace norm $tr [X^T(\sigma_y \otimes
\sigma_y ) X]$. Hence, the upper bound of LE is:
\begin{eqnarray}
E_{ij}&\leq&\frac{\sqrt{s_+^{ij}}+\sqrt{s_-^{ij}}}{2}\ ,
\label{enu}
\end{eqnarray}
where
\begin{eqnarray}
s_{\pm}^{ij}&=& \left( 1 \pm \langle\phi|
\sigma_z^{(i)}\sigma_z^{(j)} |\phi\rangle \right)^2-
\left(\langle\phi| \sigma_z^{(i)} |\phi\rangle \pm \langle \phi|
\sigma_z^{(j)}|\phi \rangle\right)^2.\nonumber
\end{eqnarray}
The lower bound on LE is expressed in terms of correlation
functions:
\begin{eqnarray}
Q_{\alpha\beta}^{ij}&=&\langle\langle\phi|\sigma_\alpha^{(i)}\sigma_\beta^{(j)}|\phi\rangle\rangle\nonumber\\
&\equiv&
\langle\phi|\sigma_\alpha^{(i)}\sigma_\beta^{(j)}|\phi\rangle
-\langle\phi|\sigma_\alpha^{(i)}|\phi\rangle\langle\phi|\sigma_\beta^{(j)}|\phi\rangle\
.
\end{eqnarray}
The lower bound on LE is  based on the following observation
\cite{vpc}: Given a state of $N$ spins with  fixed correlation
functions $Q_{\alpha\beta}^{ij}$ between two spins (marked by $i$
and $j$) and directions $\alpha$ and $\beta$, there exist
directions in which one can measure other spins [assisting spins],
such that this correlation does not decrease. Using
this  observation, VPC found a lower bound for LE:
\begin{eqnarray}
E_{ij}\ge \max_{\alpha} \left(|Q_{\alpha \alpha}^{ij}|\right)\ .
\label{enl}
\end{eqnarray}
VPC explicitly evaluated these bounds for the ground state of the
Ising model and showed that actual value of LE is close to the
lower bound.

In this paper we   consider the ground state of infinite
anti-ferromagnetic $XXX$ spin chain at zero temperature. We also consider
 anisotropic version: $XXZ$ chain.
We  calculated the concurrence before the measurement [see Appendix A]
and compare it to LE. Measurement raises the  concurrence.

\section{XXX Antiferromagnetic Spin Chain}

The Hamiltonian for anti-ferromagnetic $XXX$ spin chain can be
written as
\begin{eqnarray}
H_{XXX}^0=\sum_m \left\{ \sigma_x^{(m)}\sigma_x^{(m+1)}
+\sigma_y^{(m)}\sigma_y^{(m+1)} +\sigma_z^{(m)}\sigma_z^{(m+1)}
\right\}
\end{eqnarray}
Here $\sigma_x^{(m)}$, $\sigma_y^{(m)}$, $\sigma_z^{(m)}$ are
Pauli matrix, which describe spin operators on $m$-th lattice
site. Summation  goes through
 the whole infinite lattice.
The density of the Hamiltonian is a linear function of the swap gate.

Hans Bethe found
 eigenfunctions of the Hamiltonian of the model in 1931\cite{B}.
The ground state  $| \phi \rangle $ was found by Hulten in
Ref.~\onlinecite{H}. We shall normalize it to $1$. Correlations
are defined as
 averages with respect to the ground state. They are isotropic
\begin{eqnarray}
\langle \phi |\sigma_\alpha^{(i)}\sigma_\beta^{(j)}| \phi \rangle=
\delta^{\alpha \beta }
\langle\phi|\sigma_z^{(i)}\sigma_z^{(j)}|\phi\rangle\ .
\end{eqnarray}
There is no magnetization $\sigma_\alpha$ ($\alpha =x,y$ or $z$)
\begin{equation}
\sigma_\alpha =\langle \phi| \sigma_\alpha ^{(j)} |\phi\rangle =0\
.
\end{equation}
This simplifies the lower bound (\ref{enl}) of LE \footnote{In
this case the upper bound of $E_{ij}$ given by (\ref{enu}) is $1$
(any concurrence is bounded by $1$).}:
\begin{equation}
1\ge E_{ij}\ge |\langle \phi |\sigma_z^{(i)}\sigma_z^{(j)}| \phi
\rangle|\ .
\end{equation}
Let us now use the  explicit expression for correlations  $\langle
\phi |\sigma_z^{(i)}\sigma_z^{(j)}| \phi \rangle $ to calculate
the lower bound:
\begin{eqnarray}
\langle\phi| \sigma^{(m)}_{z}
\sigma^{(m+1)}_z|\phi\rangle&=&\frac{1}{3}-\frac{4}{3}\ln2
\simeq-0.5908629072\ , \label{1st} \\
\langle\phi| \sigma^{(m)}_{z}
\sigma^{(m+2)}_z|\phi\rangle&=&\frac{1}{3}-\frac{16}{3}\ln2+
3\zeta(3)\nonumber\\
&\simeq& 0.2427190798\ , \label{2nd}\\
 \langle\phi|
\sigma_z^{(m)} \sigma_z^{(m+3)}|\phi\rangle&=&\frac{1}{3}- 12\ln
2+\frac{74}{3}\zeta(3) -\frac{56}{3}\zeta(3)\ln 2\nonumber \\
&&-6\zeta(3)^2 -\frac{125}{6}\zeta(5)+\frac{100}{3}\zeta(5)\ln
2\nonumber\\
& \simeq & -0.2009945090\, \label{3rd}\\
\langle\phi| \sigma_z^{(m)}
\sigma_z^{(m+4)}|\phi\rangle&=&\frac{1}{12}-\frac{16}{3}\ln
2-54 \ln 2\, \zeta(3)-\nonumber\\
&&\frac{293}{4}\zeta^2(3)-\frac{875}{12} \zeta(5)+\frac{145}{6}
\zeta(3)\nonumber\\
&&+\frac{1450}{3}\ln 2\, \zeta(5)-\frac{275}{16} \zeta(3)
\zeta(5)\nonumber\\
&&-\frac{1875}{16} \zeta^2(5)+\frac{3185}{64} \zeta(7)
-\nonumber\\
&&\frac{1715}{4} \ln 2 \zeta(7) +\frac{6615}{32} \zeta(3)
\zeta(7)\nonumber\\
&\simeq& 0.0346527769. \label{4th}
\end{eqnarray}
It took a long time to evaluate correlations functions. Nearest
neighbor correlation can be extracted from the ground state
energy\cite{H}. Next to nearest neighbor correlation was
calculated by M. Takahashi  in 1977, see
Ref.~\onlinecite{Takahashi77}. Recently it was
 established that all correlations can be expressed as polynomials
 of $\ln 2$ and the  values of Riemann zeta
function \footnote{$\zeta (s) =\sum_{n=1}^\infty n^{-s}$ } at odd
arguments\cite{bk1,bk2,bkns,bks,bks2,bks3}. These polynomials have
only rational coefficients. Third neighbor correlation
 was calculated by  K. Sakai, M. Shiroishi, Y. Nishiyama, M. Takahashi ,
 see [\onlinecite{ssnt}].
These results give us the following bounds for localizable
entanglement [LE]:
\begin{eqnarray}
 E_{j,j+1} &\ge &  0.5908629072\ ,  \label{n1st}\\
E_{j,j+2}  &\ge &  0.2427190798\ ,   \label{n2nd}\\
E_{j,j+3}  &\ge &   0.2009945090\ .   \label{n3rd}\\
E_{j,j+4}  &\ge &   0.0346527769\ .   \label{n4rd}
\end{eqnarray}
At large distances correlation functions exhibits critical
behavior. Asymptotic was obtained in \cite{luk,af}:
\begin{eqnarray}
\langle\phi|\sigma_z^{(i)}\sigma_z^{(j)}|\phi\rangle \to (-1)^{i-j}
\left\{ \frac{(2 \ln |i-j|)^{1/2}}{(\pi )^{3/2}|i-j|} \right\}.\label{largexxx}
\end{eqnarray}
This helps us to estimate localizable entanglement asymptotically
for two spins, which are far away, i.e. $|i-j| \rightarrow
\infty$:
\begin{eqnarray}
E_{ij} \ge \left\{ \frac{(2 \ln |i-j|)^{1/2}}{(\pi )^{3/2}|i-j|}
\right\}\ .  \label{elargexxx}
\end{eqnarray}

Even better, but more complicate expression for lower bound of LE
, can be extracted from the paper
\cite{lt}:
 \begin{widetext}
\begin{eqnarray}
 E_{ij}\ge \sqrt{\frac{2}{\pi^3}}
    \frac{1}{|i-j| \sqrt{g}}\,
   \bigg\{\,  1+\Big(\, \frac{3}{8}-\frac{c}{2}\,\Big)\,
   g +\Big(\,\frac{5}{128}-\frac{c}{16}
          -\frac{c^2}{8}\,\Big)\, g^2 +
         \Big(\,\frac{21}{1024} +\frac{7 c}{256}
          -\frac{7 c^2}{64}
          -\frac{c^3}{16}
          +\frac{13\, \zeta(3)}{32}\,\Big)\, g^3 +O(g^4)\,
          \bigg\} \nonumber\\
          -\frac{(-1)^{|i-j|}}{\pi^2\,  |i-j|^2}\,
          \bigg\{\, 1
          +\frac{g}{2}+\Big(c+\frac{3}{4}\,\Big)\,
           \frac{g^2}{2}+\frac{c(c+2)}{2}\, g^3+O(g^4) \,
           \bigg\}+\ldots\hspace{7.0cm}
           \end{eqnarray}
\end{widetext}
Here the coupling constant $g$ depends on the distance $|i-j|$. It is
defined by:
\begin{equation}\label{cju}
   \sqrt{g}\, e^\frac{1}{g}
      = 2\sqrt{2\pi}\, e^{\gamma_E+c} |i-j|\ .
\end{equation}
Here $ \gamma_E = 0.5772 $ is the Euler's constant and  $c$ is a
parameter [normalization point]. A good choice for
$c$ is  $c=-1$.  This boundary for LE  is suitable
 for the full range of distances.
In Appendix A we calculated concurrence before the measurement. It
is nonzero only for nearest neighbors, see Ref.~\onlinecite{ch}
and (\ref{nn}). Clearly, the  measurement raises the concurrence.

\section{Critical XXZ antiferromagnet}

Let us consider effects of anisotropy of interaction of spins. The
 Hamiltonian of the XXZ spin chain is:
\begin{eqnarray}
{\bf H}_{XXZ}^0&=&-\sum_{m}\big\{ \sigma^{(m)}_{x}
\sigma^{(m+1)}_{x}+\sigma^{(m)}_{y}\sigma^{(m+1)}_{y}\nonumber\\
&&\qquad + \Delta(\sigma^{(m)}_{z}\sigma^{(m+1)}_{z}-1) \big\}\ .
\end{eqnarray}
We shall consider critical regime ($-1\leq\Delta<1 $) and we use parametrization:
\begin{equation}
\Delta=\cos(\pi\eta)\ , \qquad 0<\eta< 1\ . \label{etaf1}
\end{equation}
Let us remind that the  case $\eta=0$ corresponds to ferromagnetic XXX, which
 we are not considering here. Another case $\eta =1$  corresponds to
anti-ferromagnetic XXX, see previous section.

The case $\eta=2/3$ corresponds to  $\Delta= -1/2$. In this case
the model admits much simpler solution than generic Bethe Anzats,
see Ref.~\onlinecite{s,rs1,rs2} . In this case  the  model is
super-symmetric, see Ref.~\onlinecite{fns}.

Later we shall see that all these special cases $\eta=0, 2/3, 1$
have high sensitivity to magnetic field interacting with spins.

\begin{figure}
\includegraphics[width=3in]{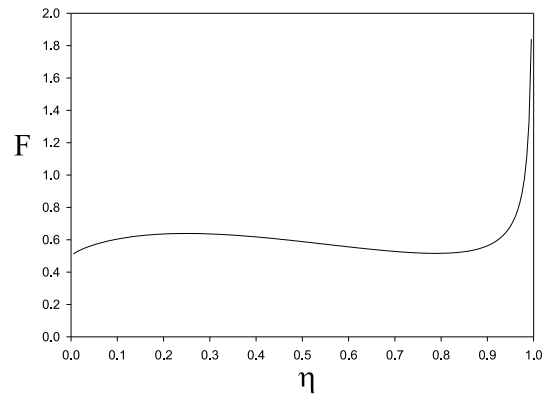}
\caption{$F$ (Eq.~\ref{Fdef}) versus $\eta$} \label{feta}
\end{figure}

For general values of $\eta$ (\ref{etaf1}) there is  no magnetization:
\begin{equation}
\sigma_\alpha =\langle \phi| \sigma_\alpha ^{(m)} |\phi\rangle =0\
.
\end{equation}
 LE is bounded by maximal correlation function:
\begin{equation}
1\ge E_{ij}\ge \max_\alpha |\langle \phi
|\sigma_\alpha^{(i)}\sigma_\alpha^{(j)}| \phi \rangle|\ .
\end{equation}
Correlation functions decays as power laws at large distances $
\ln |i-j|\gg 1/(2-2 \eta)$.  Leading terms of correlations are:
\begin{eqnarray}
\langle\phi| \sigma_x^{(i)}\sigma_x^{(j)}|\phi\rangle &=&
\langle\phi| \sigma_y^{(i)}\sigma_y^{(j)}|\phi\rangle=F
|i-j|^{-\eta}\ , \nonumber \\
\langle\phi| \sigma_z^{(i)}\sigma_z^{(j)}|\phi\rangle&=&-{1\over
 \pi^2 \eta}|i-j|^{-2}+A(-1)^{i-j}|i-j|^{-{1\over\eta}}.\nonumber
\label{zzc}
\end{eqnarray}
Many people worked on the subject. Important results are obtained
in \cite{luk}. A good collection of references can be found in the
book \onlinecite{korepin}, see pages 512, 549-553. Since $0<\eta<
1 $, it becomes clear that $\sigma_x$ correlations asymptotically
dominate the lower bound:
\begin{equation} |Q_{xx}^{ij}|=|Q_{yy}^{ij}|>|Q_{zz}^{ij}|\ .
\end{equation}
Finally we got the following bound for LE:
\begin{equation}
 E_{ij}\ge F |i-j|^{-\eta}
\end{equation}
{\it This shows that anisotropy raises the lower bound for {\rm
LE}.} The  coefficient $F$  was calculated\cite{luk,af}
\begin{eqnarray}
  F={1\over 2
(1-\eta)^2}\ \biggl[ { \Gamma\big({\eta\over 2-2\eta}\big)\over
2\sqrt\pi\ \Gamma\big({1\over 2-2\eta}\big)}
\biggr]^{\eta}\times \qquad\nonumber\\
 {\rm exp}\biggl\{- \int_{0}^{\infty} {d
t\over t}\ \Big( { {\rm sinh}\big(\eta t \big)\over {\rm sinh}( t)
{\rm cosh}\big((1-\eta) t\big )}- \eta e^{-2
t}\Big)\biggr\}.\label{Fdef}
\end{eqnarray}
The plot is shown on Fig.~\ref{feta}. There is singularity in
function $F$ for $\eta$ near $1$:
\begin{equation}
F\sim (1-\eta)^{-\frac{1}{2}}\quad \textrm{when} \quad \eta \to
1-0
\end{equation}
Special case  $\eta =1$ corresponds to XXX antiferromagnet.

In  Appendix A, we evaluated concurrence before the measurement.
It vanishes at finite distance, see (\ref{xxz}).

\section{XXX antiferromagnet in a magnetic field}

Let us come back to XXX model
 \begin{eqnarray}
H_{XXX}^0=\sum_m  \sigma_x^{(m)}\sigma_x^{(m+1)}
+\sigma_y^{(m)}\sigma_y^{(m+1)} +\sigma_z^{(m)}\sigma_z^{(m+1)}\quad
\end{eqnarray}
But now let us add magnetic field
\begin{equation}
{\bf H}_{XXX}^h=H_{XXX}^0 - \sum_{m} h \sigma^{(m)}_z\ .
\end{equation}
This introduce anisotropy in a different way. In small magnetic
field  $h \to 0$, small magnetization develops: $\sigma_z = h /
\pi^2$. For stronger magnetic field, magnetization increases. As
magnetic field approaches its critical value $h_c=4 $,
magnetization approaches $1$ [ferromagnetic state with all spins
up]:
\begin{eqnarray}
\sigma_z&=&1-\frac{2}{\pi} \sqrt{h_c-h}\ , \qquad h\to h_c-0 .
\end{eqnarray}
Exact expression for magnetization $\sigma_z$ at arbitrary value
of magnetic field is given on the pages 70-71 of the book
\onlinecite{korepin}. Averages of other spin components over
ground state are  zero: $\sigma_x=\sigma_y=0$.

In this section, we are considering moderate magnetic field $0\le
h\le h_c$. Asymptotic of correlation functions at large distances
can be described as
 follows:
\begin{eqnarray}
\langle\langle\phi| \sigma_z^{(i)}
\sigma_z^{(j)}|\phi\rangle\rangle=A_1 \frac{1}{(i-j)^2}+ A_2
\frac{\cos\{ \pi (1-\sigma_z) |i-j|\} }{|i-j|^{\theta}}\quad \label{asz}
\end{eqnarray}
Here double bracket in the left hand side means that we subtracted
$\sigma_z^2$, see (4).
 The coefficients $A_1$ and $A_2$ depend on magnetic field. For small magnetic
 field,
critical index $\theta$ is close to $1$:
\begin{eqnarray}
\theta&=&1+[2\ln(h_x/h)]^{-1}\quad h\to 0;\quad
h_x=\sqrt{\frac{8\pi^3}{e}}\label{theta2}
\end{eqnarray}
and for the values of magnetic field close to critical point,
$\theta$ is close to $2$:
\begin{eqnarray}
\theta&=&2(1-\frac{1}{\pi} \sqrt{h_c-h}),\quad h\to h_c; ~ h\le
h_c=4 .
\end{eqnarray}
In Appendix B we discuss the dependence of $\theta$ on magnetic field
for intermediate fields.  Fig. 4 for  $\eta =1$ shows that  $\theta$ is a
 monotonic function of the magnetic field.
Asymptotic of other correlation functions are:
\begin{eqnarray}
\langle\phi| \sigma^{(i)}_x\sigma^{(j)}_x|\phi\rangle
=\langle\phi| \sigma^{(i)}_y\sigma^{(j)}_y|\phi\rangle=A(h)
|i-j|^{-1/{\theta}}\label{asx}.
\end{eqnarray}
Coefficient $A(h)$ vanishes as magnetic field approaches the
critical value. Exact formula for $\theta$ at any value of
magnetic field can be found  on the pages 73-76 of the book
\onlinecite{korepin} and in \cite{luk}. It shows that $1/2 \leq
1/\theta \leq 1 \leq  \theta \leq 2 $. This means that the lower
bound of LE is dominated by $\sigma_x$ correlations again
\begin{equation}
  E_{ij}\ge A(h) |i-j|^{-1/{\theta}}\ .
\end{equation}
Now let us discuss the upper bound (3) of LE. Because of translational invariance we have:
\begin{eqnarray}
s_{+}^{ij}&=& \left( 1 + \langle\phi|
\sigma_z^{(i)}\sigma_z^{(j)}
|\phi\rangle \right)^2- 4 \sigma_z^2\nonumber\\
s_{-}^{ij}&=& \left( 1 - \langle\phi| \sigma_z^{(i)}\sigma_z^{(j)}
|\phi\rangle \right)^2\ .
\end{eqnarray}
At large space separations $|i-j| \to \infty$, correlations can be
simplified  $\langle\phi| \sigma_z^{(i)}\sigma_z^{(j)}
|\phi\rangle \to \sigma_z^2 $. This means that both $s_{\pm}^{ij}$
approach $(1-\sigma_z^2)^2$. Finally the  bounds for LE for large
$|i-j|$ are
\begin{equation}        A(h) |i-j|^{-1/{\theta}} \le
E_{ij}<1-\sigma_z^2\ .
\end{equation}
So magnetic field increases lower bound and deceases upper bound.
When magnetic field are close to the critical value, the bounds
become:
\begin{equation}
 A(h) |i-j|^{-\frac{1}{2}} \le E_{ij}<\frac{4}{\pi}\sqrt{h_c-h}\ .
\end{equation}
In  Appendix A we evaluated concurrence before the measurement.
It vanishes at finite distance, see (\ref{mag}).

In  the most general case of 
XXZ  a magnetic field
correlations can be described by the similar formulae, but parameters
$h_c$, $\sigma_z$, $\theta  $  are different. We shall elaborate in the next section.

\section{XXZ antiferromagnet in a magnetic field}

Let us add interaction with a magnetic field to XXZ
spin chain:
\begin{equation}
{\bf H}_{XXZ}^h=H_{XXZ}^0 - \sum_{m} h \sigma^{(m)}_z\ .
\end{equation}
Small magnetic field  leads to a small
magnetization
 $\sigma_z=\chi h$. The magnetic susceptibility $\chi$ is:
\begin{equation}
\chi= {1-\eta \over \pi \eta \sin \pi \eta } \label{chif1}
\end{equation}
Here we used parameter $\eta$ related to anisotropy $\Delta =\cos
\pi \eta $. The dependence of $\chi$ on $\eta$ is illustrated in
the Fig 2.
\begin{figure}
\includegraphics[width=3in]{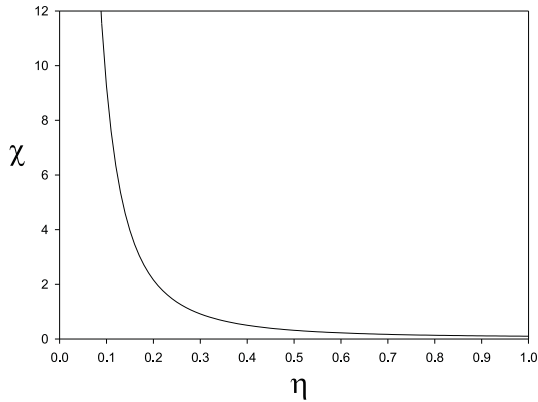}
\caption{$\chi$ (Eq.~\ref{chif1}) versus $\eta$ (Eq.~\ref{etaf1})}
\end{figure}
Let us discuss the plot. The meaning of a singularity at  $\eta=0$ is the following: The case $\eta=0$ corresponds to ferromagnetic XXX.
At zero magnetic field it has spontaneous magnetization pointed in an
 arbitrary direction.
Weak magnetic field will align spins to the  direction  of the magnetic field.
This makes susceptibility infinite.
 As $\eta$ approaches 1,
susceptibility approaches $1/\pi^2$ [antiferromagnetic XXX case].

\begin{figure}
\includegraphics[width=3in]{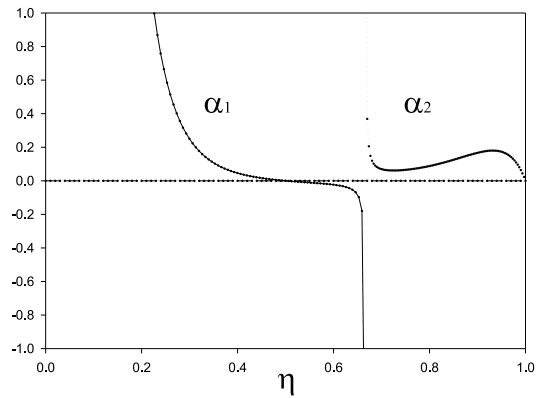}
\caption{$\alpha_1$ (Eq.~\ref{alpha1f2}), $\alpha_2$
(Eq.~\ref{alpha2f2}) versus $\eta$
(Eq.~\ref{etaf1})}\label{alphaeta}
\end{figure}

For stronger magnetic field, magnetization increases. As magnetic
field approaches its critical value $h_c=2(1-\Delta ) $,
magnetization approaches $1$:
\begin{equation}
\sigma_z=1-\frac{2}{\pi} \sqrt{h_c-h}\ , \qquad h\to h_c-0 .
\end{equation}
Averages of other spin components over ground
state are  zero: $\sigma_x=\sigma_y=0$.
Here we are considering moderate magnetic field $0\le
h\le h_c$. Asymptotic of correlation functions at large distances
can be described by the formulae similar to XXX in a magnetic field case
see (\ref{asz}), (\ref{asx}), but critical index $\theta$ is different.
A formula for  $\theta$ depends on anisotropy $\Delta =\cos \pi \eta $.
Let us first discuss small magnetic field $h\rightarrow 0$. Critical index
is quadratic in  magnetic field for $0\le \eta \le 2/3$:
\begin{equation}
\theta ={1\over \eta}\left(1+\alpha_1 h^2  \right)
\end{equation}
The coefficient
\begin{equation}
\alpha_1={(1-\eta)^2\over 4\pi \eta \tan \left( {\pi \eta \over
2(1-\eta ) } \right) \sin^2 \pi \eta } \label{alpha1f2}
\end{equation}
 shows singularity at points $\eta=0$ [ferromagnetic XXX] and
$\eta=2/3$ [$\Delta= -1/2$ case]:
\begin{eqnarray}
\alpha_1&\sim& \frac{1}{\pi^4}\frac{1}{\eta^4}
\qquad\textrm{for}\quad \eta\to 0+0\\
 \alpha_1&\sim& \frac{1}{81
\pi^2}\frac{1}{(\eta-2/3)} \qquad\textrm{for}\quad \eta \to
\left(\frac{2}{3}\right)-0
\end{eqnarray}
The nature of dependence of the coefficient $\alpha_1$ is illustrated on Fig 3.
In case  $2/3\le \eta \le 1$ small $h$ behavior is more complicated:
\begin{equation}
\theta ={1\over \eta}\left(1+\alpha_2 h^{4(\eta^{-1}-1)}  \right)
\end{equation}
Notice that the power of magnetic field changes monotonically
from 2 at $\eta=2/3$ to 0 at $\eta=1$.
An expression for the coefficient $\alpha_2 $ is more
complicated:
\begin{equation}
\alpha_2={2 \eta }e^{2\beta \over \eta} h_0^{4(1- \eta^{-1})} \tan
\left( {\pi \over \eta } \right) {\Gamma^2\left(1+{1\over \eta}
\right) \over \Gamma^2 \left({1\over 2} +{1\over \eta} \right)}
\label{alpha2f2}
\end{equation}
Here
\begin{equation}
\beta =(1-\eta) \ln (1-\eta) + \eta \ln \eta
\end{equation}
and
\begin{equation}
h_0={4 \eta \sqrt{\pi} \sin \pi \eta \over (1-\eta )}e^{\beta \over 2( 1- \eta)}
 {\Gamma\left({3-2\eta\over 2( 1- \eta)  } \right) \over \Gamma \left({2-\eta\over 2( 1- \eta)} \right)} \label{h_0}
\end{equation}
$\alpha_2(\eta)$ shows singularity at point $\eta=2/3$
\begin{eqnarray}
 \alpha_2\sim \frac{1}{81
\pi^2}\frac{1}{(\eta-2/3)} \qquad\textrm{for}\quad \eta \to
\left( \frac{2}{3}\right)+0
\end{eqnarray}
The nature of dependence of $ \alpha_2$ on $\eta$ is illustrated on
Fig.~\ref{alphaeta}.
We see that at $\eta= 2/3$ critical  index $\theta$ strongly
depends on weak magnetic field. It also depends strongly on weak
magnetic field for
 $\eta=1$, which is antiferromagnetic XXX case.

For magnetic field close to critical, the index $\theta$ approaches
2:
\begin{equation}
\theta=2+{4\sqrt{h_c-h}\over \pi \tan \left( { \pi \eta \over 2}\right) \tan{\pi \eta } }
\end{equation}
In Appendix B we discuss the dependence of $\theta$ on magnetic field
for intermediate fields. Fig. 4 shows the dependence of $\theta$ on magnetic
 field for different values of $\eta$. Note that the dependence is monotonic.

For XXZ in a magnetic field the  lower bound of localizable entanglement is
 also given by
 $\sigma_x$ correlations
\begin{equation}
E_{ij}\ge \langle\phi| \sigma^{(i)}_x\sigma^{(j)}_x|\phi\rangle
\end{equation}
Asymptotic  of the  $\sigma_x$ correlations is still given by the
formula  (\ref{asx}) with $\theta$ described in this
section.

\section{Summary}

In this paper we showed that correlations in  spin chains are
important not only for condensed mater physics and statistical
mechanics but also for quantum information. We considered
boundaries for localizable entanglement in the ground state of
antiferromagnetic spin chains. We showed that anisotropy raises
the localizable entanglement. We also calculated concurrence before the
 measurement to illustrate that the  measurement raises the concurrence.

 There are still
two open problems left: i. to prove that localizable entanglement
coincide with the lower bound $E_{ij}= \max_{\alpha}
\left(|Q_{\alpha \alpha}^{ij}|\right)$. ii. to calculate
localizable entanglement for positive  temperature.

\section*{Acknowledgments}

We are grateful to I.\ Cirac and S.\ Lukyanov for discussions. The
paper was supported by NSF Grant  DMR-0302758.

\section*{Appendix A}

In this Appendix, we calculate the concurrence between two
marked spins before measurement.
 We show that the concurrence between $i$-th and $j$-th
qubits vanishes at finite distance $|i-j|$ before the measurement.
 The density matrix
$\rho$ of $i$-th and $j$-th  spins can be represented as:
\begin{equation}
\rho=\frac{1}{4}
\sum_{\mu_i}\sum_{\mu_j}(\sigma^{(i)}_{\mu_i}\otimes
\sigma^{(j)}_{\mu_j})  \langle \sigma^{(i)}_{\mu_i}
\sigma^{(j)}_{\mu_j} \rangle
\end{equation}
To calculate concurrence we need
\begin{equation}
\tilde{\rho}=
(\sigma_y\otimes\sigma_y)\rho^*(\sigma_y\otimes\sigma_y).
\end{equation}
Here $\rho^*$ is the complex conjugate of $\rho$. Subindex  $\mu$ runs though
four different values
 $\mu=0,x,y,z$. Pauli matrix $\sigma_{\mu}$ are
\begin{eqnarray}
\sigma_0&=&I=\left( \begin{array}{cc}
1 &0\\
0& 1
\end{array}\right), \quad ~\sigma_x =\left( \begin{array}{cc}
0 &1\\
1& 0
\end{array}\right),\\
\sigma_y&=&\left( \begin{array}{cc}
0 &-i\\
i& 0
\end{array}\right), \qquad ~~\sigma_z=\left( \begin{array}{cc}
1 &0\\
0& -1
\end{array}\right).
\end{eqnarray}
Following  W.K.Wooters \cite{wot} we define
\begin{equation}
R=\sqrt{\sqrt{\rho}\,\tilde{\rho}\,\sqrt{\rho}}
\end{equation}
 We shall denote the eigenvalues of $R$
by $\lambda_k$ in a decreasing order
$$\lambda_1\ge\lambda_2\ge\lambda_3\ge\lambda_4.$$
Then concurrence ($C$) can be expressed as
$$C=\max\{0,\lambda_1-\lambda_2-\lambda_3-\lambda_4\}.$$

Let us start with the most general case $XXZ$ in a magnetic field:
\begin{eqnarray}
\langle\sigma^{(j)}_x\rangle=\langle\sigma^{(j)}_y\rangle=0;\quad \langle \sigma^{(j)}_z\rangle=\sigma \quad \textrm{and}\quad 0\le \sigma <1; \\
\langle\sigma^{(i)}_x\sigma^{(j)}_y\rangle=0;\qquad\langle\sigma^{(i)}_x\sigma^{(j)}_z\rangle=\langle\sigma^{(i)}_y\sigma^{(j)}_z\rangle=0;\\
\langle\sigma^{(i)}_x\sigma^{(j)}_x\rangle=\langle\sigma^{(i)}_y\sigma^{(j)}_y\rangle=g_x(i-j);\\
\langle\sigma^{(i)}_z\sigma^{(j)}_z\rangle=G(i-j)=\sigma^2+g_z(i-j);\\
 0<|g_z(i-j)|<|g_x(i-j)|<1\quad \mbox{for large} \quad |i-j|.
\end{eqnarray}
Hence density matrix $\rho$ for $i$-th and $j$-th qubits (spins) can be
 expressed as
\begin{eqnarray}
\rho&=&\frac{1}{4}I\otimes
I+\frac{g_x(i-j)}{4}(\sigma_x\otimes\sigma_x+\sigma_y\otimes\sigma_y)\nonumber\\
&&+\frac{G(i-j)}{4}\sigma_z\otimes\sigma_z+\frac{\sigma}{4}(I\otimes\sigma_z+\sigma_z\otimes
I).\label{rhoxxz}
\end{eqnarray}
All coefficients $\sigma$, $g_x(i-j)$, $G(i-j)$ are real, so
\begin{eqnarray}
\rho^*&=&\rho\\
\tilde{\rho}&=&(\sigma_y\otimes\sigma_y) \rho
(\sigma_y\otimes\sigma_y)
\end{eqnarray}
First three terms in Eq.~\ref{rhoxxz} commute with $\sigma_y\otimes\sigma_y$
 and the last term anti-commute.
Let us define
\begin{eqnarray}
\rho_0&=&\frac{I\otimes
I}{4}+\frac{g_x(i-j)}{4}(\sigma_x\otimes\sigma_x+\sigma_y\otimes\sigma_y)\nonumber\\
&& +\frac{G(i-j)}{4}\sigma_z\otimes\sigma_z \qquad \mbox{and} \\
m&=&\frac{\sigma}{4}(I\otimes\sigma_z+\sigma_z\otimes I).
\end{eqnarray}
Notice that $[ I\otimes\sigma_z+\sigma_z\otimes
I,\sigma_x\otimes\sigma_x+\sigma_y\otimes\sigma_y ]=0$.
So  we have
\begin{eqnarray}
\rho=\rho_0+m,\quad \tilde{\rho}=\rho_0-m, \quad [\rho_0, m ]=0,\\
~[\rho, m]=0, \qquad [ \tilde{\rho}, m ]=0,\qquad
[\rho,\tilde{\rho}]=0.
\end{eqnarray}
Now we can simplify the expression for the matrix $R$:
\begin{equation}
R=\sqrt{\sqrt{\rho}\,{\tilde{\rho}}\,\sqrt{\rho}}=\sqrt{\tilde{\rho}\,\rho}=\sqrt{\rho_0^2-m^2}.
\end{equation}
Using this representation we can diagonalize $R$. Corresponding four
 eigenvalues $\{ \lambda_k \} $ are:
\begin{eqnarray}
\left\{ \frac{1}{4}\sqrt{(1+G(i-j))^2-4
\sigma^2},\frac{1}{4}\sqrt{(1+G(i-j))^2-4
\sigma^2},\right.\nonumber\\
\left.
\frac{1-G(i-j)}{4}+\frac{g_x(i-j)}{2},\frac{1-G(i-j)}{4}-\frac{g_x(i-j)}{2}\right\}\label{eigr}
\end{eqnarray}
Now let us consider separately special cases:

\begin{flushleft}
\begin{enumerate}

\item[I.] {\it $XXX$ model with $h=0$}.

 In this case,
$\sigma=0$ and $$G(i-j)=g_x(i-j)=g_z(i-j)=g,$$ The set of
eigenvalues of $R$ becomes
\begin{eqnarray}
\{ \lambda_k \} = \left\{\frac{1+g}{4},\frac{1+g}{4}, \frac{1+g}{4},
\frac{1-3g}{4}\right\}.
\end{eqnarray}
To get am explicit expression for concurrence we need to consider
separately two cases:
\begin{enumerate}
\item[A:] $g>0$ (for $|i-j|=even$)
\begin{eqnarray}
\lambda_1=\lambda_2=\lambda_3=\frac{1+g}{4},\lambda_4=\frac{1-3g}{4}.
\end{eqnarray}
We can calculate the concurrence $C$
\begin{equation}
C=\max\{0,-\frac{1-g}{2}\}=0,
\end{equation}
since $|g|<1$. \item[B:] $g<0$ (for $|i-j|=odd$)
\begin{equation}
\lambda_1=\frac{1-3g}{4},
\lambda_2=\lambda_3=\lambda_4=\frac{1+g}{4}
\end{equation}
The concurrence
\begin{equation}
C=\max\{0, \frac{-3g-1}{4}\}
\end{equation}
 is non-zero only if $g<-\frac{1}{3}$. This
happens only for $j=i\pm 1$ with
\begin{equation}
g=\langle\sigma_j^z\sigma_{j+1}^z \rangle=\frac{1-4\ln
2}{3}\approx -0.591.
\end{equation}
Hence we have
\begin{eqnarray}
C_{j,j+1}&=&\ln 2 -\frac{1}{2}\approx 0.193, \nonumber \\
C_{j,j+k}&=&0 \quad \mbox{if} \quad  k>1. \label{nn}
\end{eqnarray}
\end{enumerate}
So, concurrence is non-zero only for nearest neighbors [for ground
state of $XXX$ model without  magnetic field]. It was first
discovered in Ref.~\onlinecite{ch}.

\item[II.] {\it  $XXZ$ model at $h=0$}.

In this case  $\sigma=0$. At  $|i-j|\to \infty $
\begin{equation} g_x(i-j)>0,\quad g_x(i-j)>|g_z(i-j)|\quad
 \nonumber\end{equation} and both
$g_x(i-j)$ and $g_z(i-j)$ decay as a function of the distance $|i-j|$.
Eigenvalues of $R$ become
\begin{eqnarray}
\lambda_1=\frac{1-g_z(i-j)}{4}+\frac{g_x(i-j)}{2};\\
\lambda_2=\lambda_3=\frac{1+g_z(i-j)}{4};\\
\lambda_4=\frac{1-g_z(i-j)}{4}-\frac{g_x(i-j)}{2}.
\end{eqnarray}
Now we can calculate the concurrence:
\begin{equation}
C=\max\{0, -\frac{1}{2}+g_x(i-j)-\frac{1}{2}g_z(i-j)\}
\end{equation}
It  vanishes at distance larger than $|i-j|_{min}$
\begin{eqnarray}
\frac{1}{2}&\simeq& g_x(|i-j|_{min})\simeq \frac{F}{|i-j|_{min}^\eta};\nonumber\\
 |i-j|_{min}&\simeq& (2F)^{\frac{1}{\eta}}.\label{xxz}
\end{eqnarray}

\item[III.] {\it  $XXX$ in a magnetic field}.

\begin{equation}
G(i-j)=\sigma^2+g_z(i-j)
\end{equation}
For large $|i-j|$ both $g_z(i-j)$ and $g_x(i-j)$ become small.
For magnetic field smaller then critical
\begin{eqnarray}
 g_z(i-j) \frac{1+\sigma^2}{(1-\sigma^2)^2} \ll 1 \nonumber \\
  |g_x(i-j)|>|g_z(i-j)|.
\end{eqnarray}
 Eigenvalues of
$R$ in Eq.~\ref{eigr} become
\begin{eqnarray}
\lambda_1=\frac{1-\sigma^2}{4}-\frac{g_z(i-j)}{4}+|\frac{g_x(i-j)}{2}|;\\
\lambda_2=\lambda_3=\frac{1-\sigma^2}{4}+\frac{g_z(i-j)}{4}
\frac{1+\sigma^2}{1-\sigma^2};\\
\lambda_4=\frac{1-\sigma^2}{4}-\frac{g_z(i-j)}{4}-|\frac{g_x(i-j)}{2}|,
\end{eqnarray}
and concurrence becomes
\begin{equation}
C=\max\{0,
|g_x(i-j)|-\frac{1-\sigma^2}{2}-\frac{g_z(i-j)}{2}\frac{1+\sigma^2}{1-\sigma^2}\}.
\end{equation}
Hence concurrence vanishes at distance lager than $|i-j|_{min}$
\begin{eqnarray}
\frac{1-\sigma^2}{2}&=& |g_x(|i-j|_{min})|= |A(h)| |i-j|_{min}^{-1/\theta}\nonumber \\
 |i-j|_{min}&=&|\frac{2A(h)}{1-\sigma^2}|^{\theta}.\label{mag}
\end{eqnarray}

\end{enumerate}
\end{flushleft}

\section*{Appendix B}

In this Appendix, we discuss the dependence of  critical
exponent $\theta$ on the  magnetic field $h$.
 We follow the book \onlinecite{korepin}.

\begin{flushleft}
\begin{enumerate}

\item[I.] Let us start from {\it  $XXX$ model}.

Energy of a magnon $\epsilon(\lambda)$
is defined by a set of equations:
\begin{eqnarray}
\epsilon(\lambda)&-&\frac{1}{2\pi}\int_{-\Lambda}^{\Lambda}
K(\lambda, \mu) \epsilon(\mu) \mathrm{d} \mu
=\epsilon_0(\lambda),\label{ie} \\
K(\lambda, \mu)&=&\frac{-2}{1+(\lambda-\mu)^2},\quad
\epsilon_0(\lambda)=2h-\frac{2}{\frac{1}{4}+\lambda^2}\label{xlambda}
\end{eqnarray}
With extra condition $\epsilon(\pm \Lambda)=0$.
This set of equation determines the dependence of  $\Lambda$
on magnetic field $h$. Here  $\Lambda$ is a value of a spectral
parameter at the Fermi edge.
An important object is the  fractional
charge $Z(\lambda)$:
\begin{equation}
Z(\lambda)-\frac{1}{2\pi} \int_{-\Lambda}^{\Lambda} K(\lambda,
\mu) Z(\mu) \mathrm{d} \mu=1.
\end{equation}
The critical exponent is equal to:
\begin{equation}
\theta=2 Z^2(\Lambda)
\end{equation}
For $XXX$ model, the critical field $h_c=4$. For large magnetic field
 ($|h|>h_c$) $ \Lambda=0.$ If magnetic filed approaches the critical value from below, then
\begin{equation}
 \Lambda=\frac{1}{2}\sqrt{h_c-|h|} \quad\textrm{and}\quad \theta=2-\frac{4}{\pi}
 \Lambda\to 2.
\end{equation}
If the magnetic field is small $|h|\to 0$ then
\begin{equation}
\Lambda=\frac{1}{2\pi} \ln \left(\frac{(2\pi)^3}{e h^2}\right)\to
\infty \quad\textrm{and}\quad \theta=1+\frac{1}{2\pi \Lambda}\to 1
\end{equation}

\item[II.] Now let us discuss {\it  $XXZ$ model}.

In this case  we can use the same  set of equations (\ref{ie},\ref{xlambda})
with $K(\lambda, \mu)$ and $\epsilon_0(\lambda)$ replaced by:
\begin{eqnarray}
K(\lambda, \mu)=\frac{\sin(2\pi \eta)}{\sinh (\lambda-\mu +i\pi\eta)\sinh (\lambda-\mu -i\pi\eta)},\nonumber\\
\epsilon_0(\lambda)=2h-\frac{2\sin^2(\pi \eta)}{\cosh
(\lambda+\frac{i\pi\eta}{2})\cosh (\lambda-\frac{i\pi\eta}{2})}.
\end{eqnarray}
For small magnetic field
\begin{eqnarray}
 \Lambda=(1-\eta)\ln
\left(\frac{h_0}{ h}\right) \to \infty \quad
\textrm{when}\quad h\to 0
\end{eqnarray}
Here $h_0$ is given by (\ref{h_0}).

But for magnetic field close to critical:
\begin{equation}
\Lambda=\frac{\sqrt{h_c-|h|}}{2 \tan (\frac{\pi \eta }{2})}\to 0
\quad\textrm{when}\quad h\to \pm h_c.
\end{equation}
The critical value of the magnetic field is:
\begin{equation}
h_c=2(1-\Delta)=2(1-\cos(\pi\eta))
\end{equation}
For general magnetic field $h$, we solved these equations
numerically  and found that both $\Lambda(h)$ and $\theta(h)$ are
monotonic functions of $h$.  The numerical solution for
$\theta(h)$ was shown in Fig 4.
\begin{figure}
\includegraphics[width=3in]{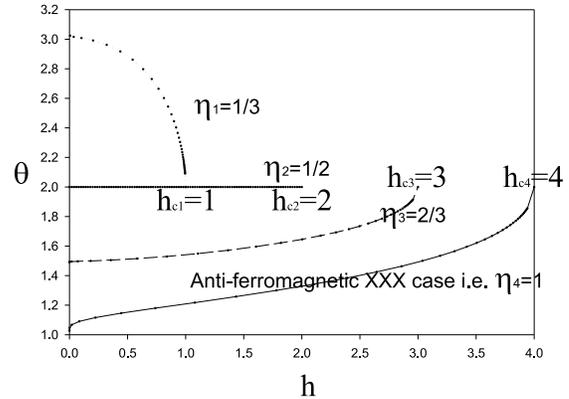}
\caption{Critical exponent $\theta$ (Eqs.~\ref{asz} and \ref{asx})
versus magnetic field $h$ for different values of anisotropy $\eta$.}
\end{figure}

\end{enumerate}
\end{flushleft}

\end{document}